# Detrimental Effect of Interfacial Dzyaloshinskii-Moriya Interaction on Perpendicular Spin-Transfer-Torque Magnetic Random Access Memory

Peong-Hwa Jang[1], Kyungmi Song[2], Seung-Jae Lee[2], Seo-Won Lee[1,†], and Kyung-Jin Lee[1,2,†]

[1]Department of Materials Science and Engineering, Korea University, Seoul 136-713, Korea

[2]KU-KIST Graduate School of Converging Science and Technology, Korea University, Seoul 136-713, Korea

[†]Correspondence to: K.-J.L. (kj_lee@korea.ac.kr) or S.-W.L. (swlee-sci@korea.ac.kr)

**Interfacial Dzyaloshinskii-Moriya interaction in ferromagnet/heavy metal bilayers is recently of considerable interest as it offers an efficient control of domain walls and the stabilization of magnetic skyrmions. However, its effect on the performance of perpendicular spin transfer torque memory has not been explored yet. We show based on numerical studies that the interfacial Dzyaloshinskii-Moriya interaction decreases the thermal energy barrier while increases the switching current. As high thermal energy barrier as well as low switching current is required for the commercialization of spin torque memory, our results suggest that the interfacial Dzyaloshinskii-Moriya interaction should be minimized for spin torque memory applications.**

Quantum mechanical interatomic magnetic exchange is decomposed into symmetric and antisymmetric components. The symmetric exchange stabilizes spatially uniform magnetization whereas the antisymmetric exchange, called Dzyaloshinskii-Moriya interaction (DMI) [1, 2], favors chiral magnetic textures. The DMI between two atomic spins $\mathbf{S}_i$ and $\mathbf{S}_j$ is given as

$$H_{DM} = -\mathbf{D}_{ij} \cdot (\mathbf{S}_i \times \mathbf{S}_j), \tag{1}$$

where $\mathbf{D}_{ij}$ is the DM vector, which is perpendicular to both the asymmetry direction and the position vector $\mathbf{r}_{ij}$ between the spins $\mathbf{S}_i$ and $\mathbf{S}_j$. In bilayers consisting of ferromagnet (FM) and heavy metal (HM), the DM vector lies in the film plane (i.e., x-y plane), which results in the interfacial DMI [3]. A considerable attention is currently being paid to the interfacial DMI as it assists fast motion of a magnetic domain wall [4-7] and stabilization of an isolated magnetic skyrmion [8, 9]. Various methodologies to estimate DMI in bilayers are also proposed, such as the anisotropic expansion of a perpendicular domain wall by an in-plane magnetic field [10] and the non-reciprocal propagation of spin waves in the presence of interfacial DMI [11-13]. Based on the latter method, the DMI value D larger than 1 erg/cm$^2$ was recently reported for Pt/Co/AlO$_x$ structures [14].

On the other hand, much efforts have been devoted to the development of perpendicular spin-transfer torque magnetic random access memories (STT-MRAMs) due to their excellent endurance and low power consumption [15]. In order to commercialize perpendicular STT-MRAMs, a small switching current density as well as a large energy barrier is required. The freely switchable layer of perpendicular STT-MRAM is commonly composed of a CoFe-based FM in contact to a HM. At the interface of FM/HM, the following three pre-conditions for the emergence of interfacial DMI are naturally satisfied; i.e., inversion symmetry breaking at the interface, strong spin-orbit coupling from HM, and exchange coupling from FM. In this respect, an important question is whether or not the interfacial DMI is beneficial for perpendicular STT-MRAMs.

In this work, we report numerical results on the effect of interfacial DMI on thermal energy barrier and switching current density of circular shaped perpendicular STT-MRAM. The energy barrier $E_B$ is calculated via the string method [16]. We initially set a transition path between two minima (approximately all up or down spins) and make discrete images of the path (image number $i = 0, \ldots, 100$). As the initial path is not an ultimate minimum energy path (MEP), we modify the path by lowering the energy of each $i^{\text{th}}$ image via $\hat{\mathbf{m}}_i(t + \Delta t) = \hat{\mathbf{m}}_i(t) - \int_t^{t+\Delta t} \gamma \alpha \hat{\mathbf{m}} \times (\hat{\mathbf{m}} \times \mathbf{H}) dt$ until reaching the MEP, where $\gamma$ is the gyromagnetic ratio, $\alpha$ (= 1, i.e., overdamping) is the damping constant, and $\mathbf{H}$ is the effective field including the exchange, DMI, magnetostatic, and anisotropy fields. We also perform the reparametrization at every 100 iteration steps, in order to keep the images equidistant in the phase space.

The switching current density is calculated by solving the stochastic Landau-Lifshitz-Gilbert equation including the Slonczewski's STT term, given as

$$\frac{d\hat{\mathbf{m}}}{dt} = -\gamma \hat{\mathbf{m}} \times \mathbf{H}_{eff} + \alpha \hat{\mathbf{m}} \times \frac{d\hat{\mathbf{m}}}{dt} + \gamma \frac{\hbar}{2e} \frac{\eta}{M_s t_F} J \hat{\mathbf{m}} \times (\hat{\mathbf{m}} \times \hat{\mathbf{z}}), \qquad (2)$$

where $\hat{\mathbf{m}}$ is the unit vector along the magnetization, $\mathbf{H}_{eff}$ is the effective field including the exchange, DMI, magnetostatic, anisotropy, thermal fluctuation, and current-induced Oersted fields, $\alpha$ (= 0.01) is the damping constant, $\hat{\mathbf{z}}$ is the unit vector along the thickness direction, $M_S$ (= 1000 emu/cm$^3$) is the saturation magnetization, $t_F$ (= 1 nm) is the thickness of free layer, $\eta$ (= 0.5) is the spin polarization factor, and $J$ is the current density. The switching current density $J_{SW}$ is determined by the current density corresponding to the switching probability of 50 % out of 30 trials for each case at the temperature of 300 K. We use the following parameters: the exchange stiffness constant $A_{ex}$ of $10^{-6}$ erg/cm, the perpendicular anisotropy energy density $K_u$ of $10^7$ erg/cm$^3$, the DMI constant D of 0~1 erg/cm$^2$, and MRAM cell diameter of 10~40 nm. We use a constant $M_S t_F$ of $10^{-4}$ erg/(Oe·cm$^2$) for various DMI constants assuming that FM is fixed whereas HM would vary. In all calculations, we consider the boundary condition caused by DMI [17]. We below show calculation results obtained from two-dimensional (2D) micromagnetic model where the free layer is not discretized in the thickness direction. We confirm that this 2D model gives almost identical results with 3D model (not shown) because the film thickness is thin enough (= 1 nm) and much smaller than the exchange length (~ 5 nm in this work).

We first show the effect of DMI on the energy barrier $E_B$ (Fig. 1). We find that the DMI always lowers the energy barrier. In Figs. 1(a) and (b), the energy at the image number of 50 corresponds to the saddle point and thus the energy barrier $E_B$. The DMI-induced decrease of energy barrier becomes more pronounced for a larger cell (Fig. 1(c)). This behavior is understood by the domain wall energy for a large cell because the switching is mediated by the domain wall formation and propagation in this case (see Figs. 1(e) and (h); cell diameter = 30 nm). The domain wall for D = 0 is of Bloch type (Fig. 1(e)), whereas the domain wall for nonzero D is of Néel type (Fig. 1(h)). When the domain wall is formed during the switching, the energy barrier is approximately given by $\gamma_{DW} W t_F$ where W is the MRAM cell diameter and $\gamma_{DW}$ is the domain wall energy per unit area given by [18]

$$\gamma_{DW} = 4\sqrt{A_{ex} K_{eff}} - \pi |D|. \tag{3}$$

Here $K_{eff}$ (= $K_u - N_d M_s^2/2$) is the effective perpendicular anisotropy energy density and $N_d$ is the out-of-plane demagnetization factor. The domain wall energy formulation naturally explains the dependence of energy barrier on the DMI strength and cell size for a cell larger than 20 nm in Fig. 1. On the other hand, we do not observe a clear domain wall formation for the cells equal to or smaller than 15 nm. In these cases, the decreased energy barrier due to DMI is mainly related with the distortion of magnetic texture at the equilibrium. For zero DMI, the equilibrium magnetizations are almost perfectly aligned along the film normal (Figs. 1(d) and (f)). For a nonzero DMI, however, the magnetizations especially at the cell edge deviates from the film normal (Figs. 1(g) and (i)). An in-plane component of magnetization at the cell edge is given by $\pm D\sqrt{K_{eff}/(4A_{ex})}$ [17] so that the deviation at the cell edge is a consequence of interfacial DMI. This deviation is responsible for the reduced energy barrier for small cells even without the domain wall formation.

One may expect that this deviation at the cell edge reduces the switching current as the STT ($\sim \hat{\mathbf{m}} \times (\hat{\mathbf{m}} \times \hat{\mathbf{z}})$) is large at the cell edge even in the initial time stage. In contrast to this naïve expectation, however, we find that the switching current increases with increasing DMI, as we will show below. Figures 2(a) and (b) show the switching probability $P_{SW}$ as a function of the current density for the cell sizes of 10 nm and 30 nm, respectively. The calculations have been done for a finite temperature of 300 K and a current pulse width of 10 ns with rise/fall time of 1 ns. They clearly show that a larger current is required to switch the magnetization as the DMI increases. The switching current densities $J_{SW}$'s for various DMI values and MRAM cell sizes are summarized in Fig. 2(c). The increased switching current density due to DMI becomes more pronounced for a larger cell. Figure 2(d) shows the switching current density $J_{SW}$ as a function of current pulse width for various DMI constants. We find that even at a very short pulse of 1 ns, DMI increases the switching current density. Therefore, the increased $J_{SW}$ due to DMI is a general feature regardless of the current pulse width.

This unexpected DMI-induced increase of switching current demands further explanation in detail. We first describe qualitative difference in magnetization dynamics between zero and nonzero D. Figures 3(a) and 3(b) respectively show temporal evolutions of the normalized average z-component of magnetization $<M_z>/M_S$ for $D = 0$ and $D = 1$ erg/cm$^2$ calculated at the same current density, $J = 9.1\times10^6$ A/cm$^2$ (T = 0 K and MRAM cell diameter = 20 nm). In the initial time stage, $<M_z>/M_S$ is slightly smaller than 1 for nonzero D (as indicated by a red up-arrow in Fig. 3(b)) while it is almost 1 for zero D. It is because for nonzero DMI, the magnetizations at the cell edge initially tilt from the easy-axis (i.e., z-axis). This initial tilting results in more efficient spin transfer in case of nonzero DMI so that a large-angle precession starts just after the current is turned on. As a result, the magnetization attempts to switch quicker for nonzero D than for zero D (i.e., the first switching trial occurs at t ~ 25 ns for $D = 1$ erg/cm$^2$ whereas it occurs at ~ 30 ns for $D = 0$). In case of zero D, this first switching trial eventually leads to the complete switching because the overall effective damping is negative and the magnetization is highly coherent (see Fig. 3(c) corresponding to "c" in Fig. 3(a)). On the other hand, in case of nonzero D, the first switching trial fails and then $<M_z>/M_S$ almost restores its initial value, leading to the oscillation of $<M_z>/M_S$ between ~0.95 and ~0.8.

We next explain the reason of the substantially different current-induced magnetization dynamics between cases of zero and nonzero DMI. In case of zero DMI, magnetizations show a collective coherent motion in the initial precession regime (Fig. 3(c)). In this case, the overall effect STT (i.e., anti-damping effect) is coherently added up for all individual local magnetizations, leading to a complete switching for a current above a certain threshold. This statement is valid even when the switching is governed by the domain wall nucleation and propagation. Figure 3(a) shows such a case where the exchange energy ($E_{exch}$) increases during the switching due to the formation of domain wall. Once a domain wall is nucleated at an edge of MRAM cell, it propagates to the other edge straightforwardly because the magnetic configuration is still coherent except the domain wall part and the STT effect makes a unidirectional motion of domain wall.

However, the magnetization dynamics is significantly different in case of nonzero DMI. As mentioned above, nonzero DMI results in a flower-like magnetic texture even in the initial equilibrium state. This DMI-induced noncollinear magnetic texture causes inhomogeneous STT effects on local magnetizations from the beginning, which leads to strongly non-uniform and time-dependent magnetic textures when STT begins to excite large-angle magnetic precession (Fig. 3(e)). These non-uniform textures inhibit the completion of switching because the transfer of angular momentum from the current becomes somewhat frustrated, with the torque applied to different regions of the sample no longer adding coherently. At the time that the magnetic texture arrives at such highly non-uniform state, the total magnetic energy is high as shown in the lower panel of Fig. 3(b). Given that the overall STT effect is weak at this moment and thus unable to further reduce $<M_z>/M_S$, the magnetic system tends to come back to its initial state (i.e. a near energy minimum state) by relaxing the energy. In Fig. 3(b), we show that this DMI-induced frustration (i.e., the reduction in $<M_z>/M_S$ corresponding "e") and energy relaxation (i.e., the recovery of $<M_z>/M_S$ back to ~1 corresponding "d") processes occur alternately, which results in the switching failure in case of nonzero DMI. Therefore, we attribute the increased switching current with increasing DMI to this unusual current-induced magnetization dynamics in the presence of DMI.

To summarize, we show that the DMI decreases the energy barrier while increases the switching current. The decreased energy barrier is caused by DMI-induced reduction in the domain wall energy. On the other hand, the increased switching current originates from the fact that the DMI combined with STT promotes strongly non-uniform magnetic textures, which make overall spin transfer effect less efficient. Our results suggest that the DMI should be minimized as high thermal energy barrier as well as low switching current is key to the commercialization of STT-MRAMs. A recent progress in theoretical studies on the microscopic origin of DMI [19-21] may help material engineering for minimizing the DMI.

This work was supported by the National Research Foundation of Korea (NRF) (NRF-2013R1A2A2A01013188, 2011-0028163), the MEST Pioneer Research Center Program (2011-0027905), and Samsung electronics.

**Figure legends**

Figure 1. Effect of interfacial DMI on energy barrier: (a) cell diameter = 10 nm, and (b) cell diameter = 30 nm. (c) The energy barrier $E_B$ normalized by $E_B$ (D = 0) at various DMI constants and cell sizes. (d-f) Magnetization configurations for D = 0: (d) image number = 0 corresponding to a local minimum (i.e., spins are almost up), (e) image number = 50 corresponding to the saddle point (i.e., $E_B$), and (f) image number = 100 corresponding to another local minimum (i.e., spins are almost down). (g-i) Magnetization configurations for D = 1 erg/cm$^2$: (d) image number = 0, (e) image number = 50, and (f) image number = 100. For (d)-(i), the cell diameter is 30 nm. The color code corresponds to the out-of-plane component of magnetization ($M_z$), and the arrows correspond to the in-plane component of magnetization.

Figure 2. Effect of interfacial DMI on switching current density: (a) cell diameter = 10 nm, and (b) cell diameter = 30 nm. (c) The switching current density $J_{SW}$ at various DMI constants and cell sizes (d) The switching current density $J_{SW}$ as a function of current pulse width at various DMI constants for the cell diameter is 20 nm.

Figure 3. Effect of interfacial DMI on switching mode (cell diameter = 20 nm, current density = 9.1×10$^6$ A/cm$^2$, T = 0 K). (a), (b) Temporal evolution of the normalized average out-of-plane component of magnetization $<M_z>/M_s$ (upper panel) and corresponding magnetic energies (lower panel) for D = 0 and D = 1 erg/cm$^2$ respectively. Here $E_{tot}$, $E_{exch}$, $E_{aniso}$, $E_{demag}$, and $E_{DMI}$ are respectively total, exchange, anisotropy, magnetostatic, and DMI energies. (c) Coherent

large-angle precession for D = 0 corresponding to "c" indicated in (a). (d) Magnetization configuration corresponding to "d" indicated in (b). (e) Magnetization configuration (incoherent large-angle precession) corresponding to "e" indicated in (b). In (c), (d), and (e), the arrows correspond to the in-plane component of magnetization and the color code corresponds to the out-of-plane component.

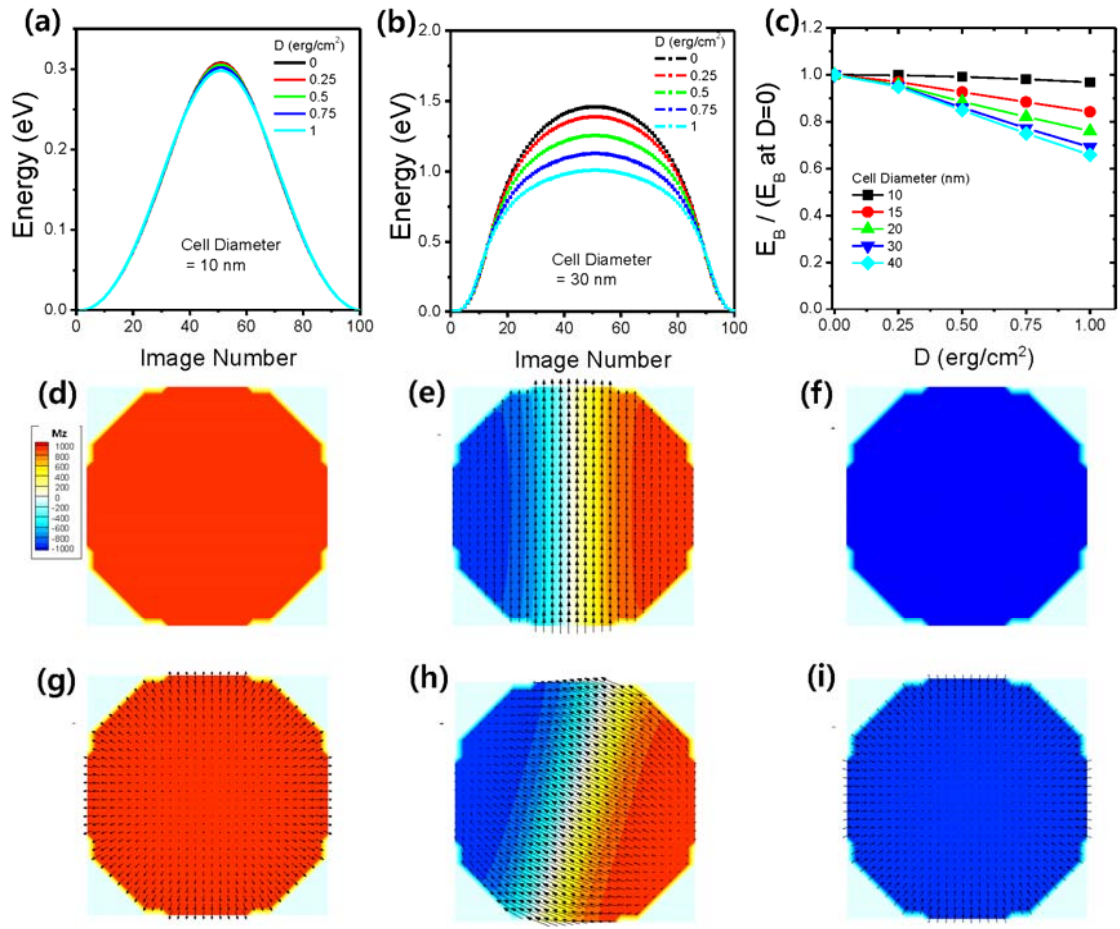

Fig. 1

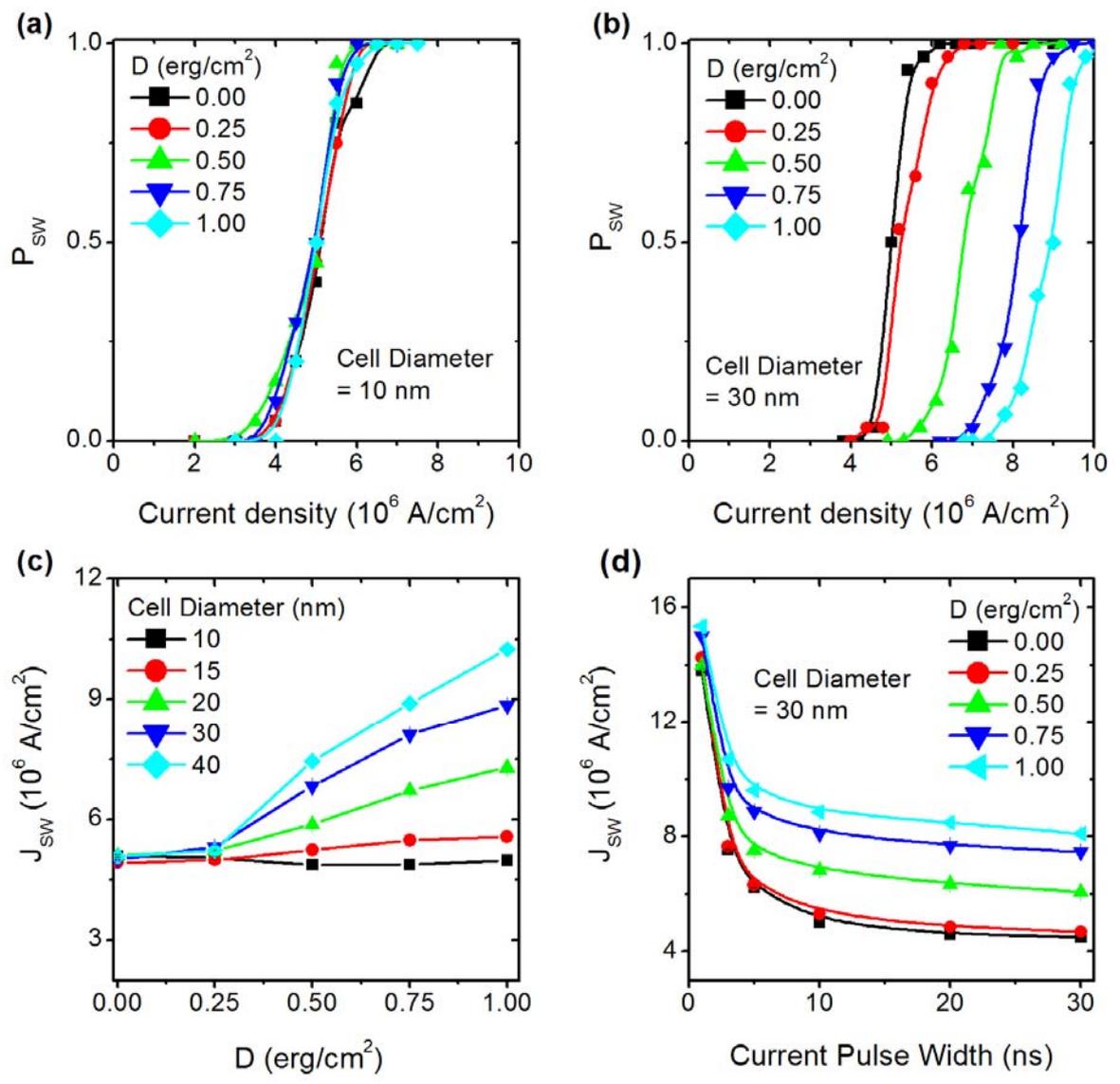

Fig. 2

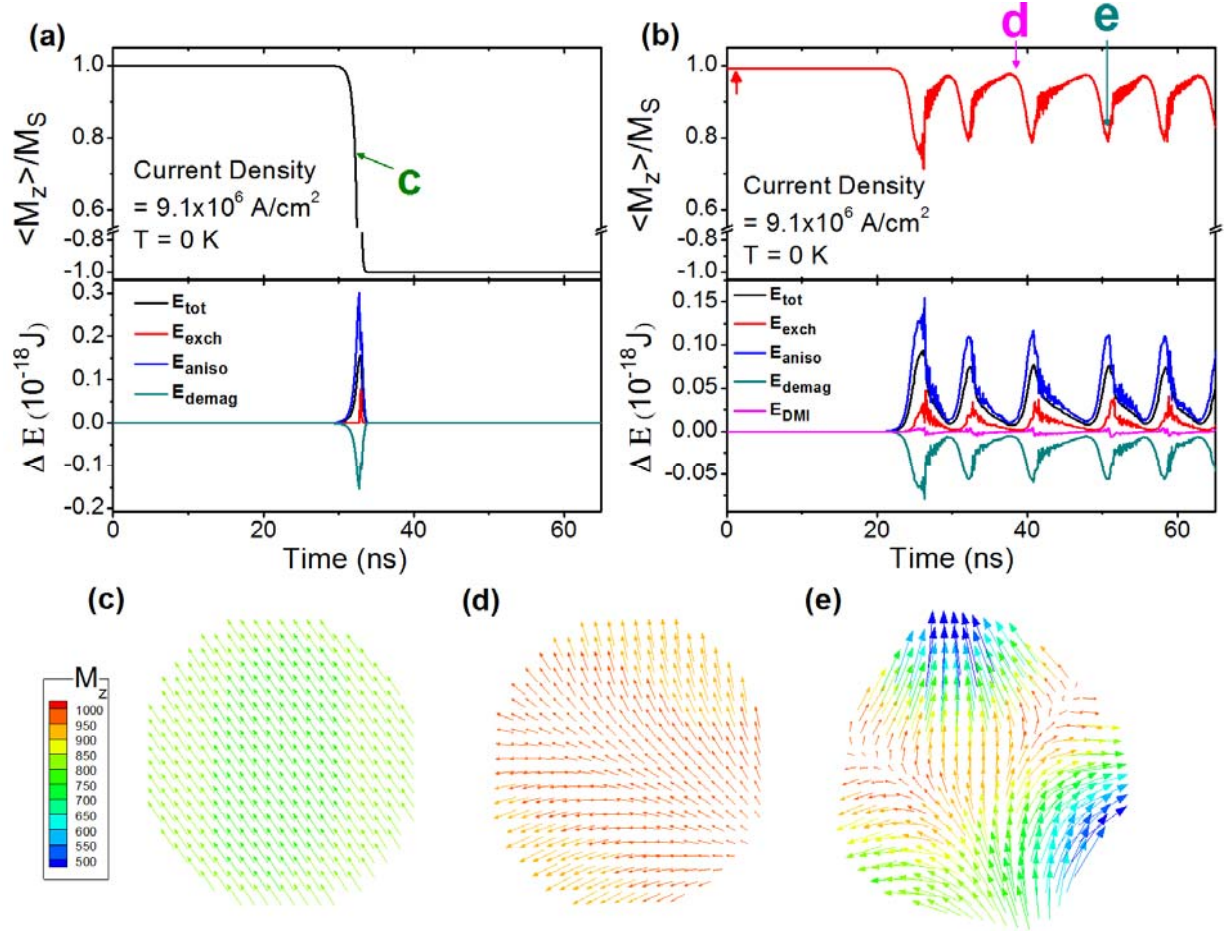

Fig. 3